\documentclass{pasj01}
\usepackage{url}
\Received{$\langle$reception date$\rangle$}
\Accepted{$\langle$acception date$\rangle$}
\Published{$\langle$publication date$\rangle$}

\RequirePackage{color}

\begin{document}

\title{Searches for New Milky Way Satellites from the First Two Years
of Data of the Subaru/Hyper Suprime-Cam Survey: Discovery of Cetus~III}

\author{Daisuke~Homma\altaffilmark{1}, Masashi~Chiba\altaffilmark{1}, 
Sakurako~Okamoto\altaffilmark{2}, Yutaka~Komiyama\altaffilmark{3,4}, 
Masayuki~Tanaka\altaffilmark{3}, Mikito~Tanaka\altaffilmark{1},
Miho~N.~Ishigaki\altaffilmark{5}, Kohei~Hayashi\altaffilmark{6,5},
Nobuo~Arimoto\altaffilmark{7,4,10},
Jos\'e~A.~Garmilla\altaffilmark{8}, Robert~H.~Lupton\altaffilmark{8},
Michael~A.~Strauss\altaffilmark{8}, 
Satoshi~Miyazaki\altaffilmark{3,4},
Shiang-Yu~Wang\altaffilmark{9}, and
Hitoshi~Murayama\altaffilmark{5}
}

\altaffiltext{1}{Astronomical Institute, Tohoku University, Aoba-ku,
Sendai 980-8578, Japan}
\altaffiltext{2}{Shanghai Astronomical Observatory, 80 Nandan Road, Shanghai 200030, China}
\altaffiltext{3}{National Astronomical Observatory of Japan, 2-21-1 Osawa, Mitaka,
Tokyo 181-8588, Japan}
\altaffiltext{4}{The Graduate University for Advanced Studies, Osawa 2-21-1, Mitaka, Tokyo 181-8588, Japan}
\altaffiltext{5}{Kavli Institute for the Physics and Mathematics of the Universe (WPI),
The University of Tokyo, Kashiwa, Chiba 277-8583, Japan}
\altaffiltext{6}{Kavli Institute for Astronomy and Astrophysics, Peking University,
Beijing 100871, China}
\altaffiltext{7}{Subaru Telescope, National Astronomical Observatory of Japan, 650 North A'ohoku Place,
Hilo, HI 96720, USA}
\altaffiltext{8}{Princeton University Observatory, Peyton Hall, Princeton, NJ 08544, USA}
\altaffiltext{9}{Institute of Astronomy and Astrophysics, Academia Sinica, Taipei, 10617, Taiwan}
\altaffiltext{10}{Astronomy Program, Department of Physics and Astronomy,
Seoul National University, 599 Gwanak-ro, Gwanak-gu, Seoul, 151-742, Korea}
\email{chiba@astr.tohoku.ac.jp}

%\keywords{galaxies: dwarf --- galaxies: individual (Cetus, Virgo) --- Local Group}

\maketitle

%%%%%% Abstract %%%%%%%%%%%%%%%%%%%%%%%%%%%%%%%%%%%%%%%%%%
\begin{abstract}
We present the results from a search for new Milky Way (MW) satellites from the first
two years of data from the Hyper Suprime-Cam (HSC) Subaru Strategic Program (SSP)
$\sim 300$~deg$^2$ and report the discovery of a highly compelling ultra-faint
dwarf galaxy candidate in Cetus. This is the second
ultra-faint dwarf we have discovered after Virgo~I reported in our previous paper.
This satellite, Cetus~III, has been identified as a
statistically significant (10.7$\sigma$) spatial overdensity of star-like objects,
which are selected from a relevant isochrone filter designed for a metal-poor and
old stellar population. This stellar system is located at a heliocentric distance
of 251$^{+24}_{-11}$~kpc with a most likely absolute magnitude of
$M_V = -2.4 \pm 0.6$~mag estimated from a Monte Carlo analysis. Cetus~III is extended
with a half-light radius of $r_h = 90^{+42}_{-17}$~pc, suggesting that this is a faint dwarf
satellite in the MW located beyond the detection limit of the Sloan Digital Sky Survey.
Further spectroscopic studies are needed to assess the nature of this stellar system.
We also revisit and update the parameters for Virgo~I finding $M_V = -0.33^{+0.75}_{-0.87}$~mag
and $r_h = 47^{+19}_{-13}$~pc.
Using simulations of $\Lambda$-dominated cold dark matter models, we
predict that we should find one or two new MW satellites from $\sim 300$~deg$^2$ HSC-SSP
data, in rough agreement with the discovery rate so far.
The further survey and completion of HSC-SSP over $\sim 1,400$~deg$^2$ will provide
robust insights into the missing satellites problem.
\end{abstract}
%%%%%%%%%%%%%%%%%%%%%%%%%%%%%%%%%%%%%%%%%%%%%%%%%%%%%%%%%%

%\keywords{galaxies: dwarf --- galaxies: individual (Cetus, Virgo) --- Local Group}

%%% Sec.1 %%%%%%%%%%%%%%%%%%%%%%%%%%%%%%%%%%%%%%%%%%%%%%%%
\section{Introduction}

The current standard theory of structure formation based on $\Lambda$-dominated
cold dark matter ($\Lambda$CDM) models is successful for understanding the origin
and evolution of observed large-scale structures in the universe on scales larger
than $\sim 1$~Mpc (e.g., \cite{Tegmark2004}). The theory states that
larger dark halos are formed by the hierarchical assembly of
smaller halos, where the latter formed earlier and thus have higher internal
densities than the former. As a consequence, dark halos like those associated with
the Milky Way (MW) are surrounded by hundreds of smaller subhalos, which survive
the merging and tidal processes \citep{Klypin1999,Moore1999}.
However, this prediction is in conflict with the observed number of only
$\sim 50$ MW satellites. This is the so-called missing satellites problem.
The theory also has several unsolved issues on small scales, including
the core/cusp problem (e.g., \cite{Moore1994,Burkert1995,de Blok2001,Swaters2003,
Gilmore2007,Oh2011}), the too-big-to-fail problem \citep{Boylan-Kolchin2011,
Boylan-Kolchin2012}, and the observed anisotropic distribution
of MW and Andromeda satellites (e.g., \cite{Kroupa2005,McConnachie2006,Ibata2013,
Pawlowski2012,Pawlowski2013,Pawlowski2015}).

A key in these small-scale issues of dark matter is to understand the basic properties
of dwarf spheroidal galaxies (dSphs) as companions of the MW and Andromeda.
DSphs are faint, metal-poor, old stellar systems, similar to globular clusters,
but they are extended and largely dominated by dark matter, unlike globular
clusters (e.g., \citep{Gilmore2007,Simon2007}). Thus, dSphs play an important role as
tracers of background dark matter, and their total number and spatial distribution in the
MW as well as their internal density profiles set invaluable constraints on
the nature of dark matter on small scales and the resultant effects on the
star-formation history of dSphs (e.g., \cite{Milosavljevic2014,Okayasu2016}).

One of the possible solutions to the missing satellites problem is that we still are
undercounting the population of dwarf satellites in the MW due to various observational
biases \citep{Koposov2008,Tollerud2008}. In particular, searches for new dSphs are
generally limited in survey area and depth. To overcome this limitation, several
survey programs have been undertaken to find new dwarf satellites in the MW,
including the Sloan Digital Sky Survey (SDSS) \citep{York2000},
the Dark Energy Survey (DES) \citep{Abbott2016}, the Pan-STARRS~1 (PS1) 3$\pi$ survey
\citep{Chambers2016}, which have revealed a number of
new ultra-faint dwarf galaxies (UFDs) with $V$-band absolute magnitude, 
$M_V$, fainter than $-8$ mag 
(e.g., \cite{Willman2005,Sakamoto2006,Belokurov2006,
Laevens2014,Kim2015,KimJerjen2015,Laevens2015a,Laevens2015b,
Bechtol2015,Koposov2015,Drlica-Wagner2015}).
It is thus expected that still more dwarf satellites remain undetected
in the outskirts of the MW because of their faint magnitudes and large distances.

This paper presents our second discovery of a new faint dwarf satellite in the MW
from the ongoing Subaru Strategic Program (SSP) using Hyper Suprime-Cam (HSC)
(see for the details of HSC-SSP, \cite{Aihara2017a,Aihara2017b}). HSC is a prime-focus
camera on Subaru with a 1.5~deg diameter field of view \citep{Miyazaki2012,Miyazaki2017}.
Our team has already reported the discovery of a new UFD, Virgo~I,
from the early data of HSC-SSP \citep{Homma2016}. In this paper, we refine and update
our method for the search of new dwarf satellites based on the analysis of statistically
significant spatial overdensities from HSC-SSP data, as presented in Section 2.
Section 3 is devoted to the results of our algorithm for detecting new satellites,
which reveals a new candidate in the direction of Cetus, named Cetus~III.
The structural parameters of Cetus~III as well as its heliocentric
distance and $V$-band absolute magnitude are also derived. In Section 4, we examine
the significance of having found these two UFDs from the early HSC data and
predict how many more UFDs would be found in the survey program.
Finally, our conclusions are presented in Section 5.

%%% Sec.2 %%%%%%%%%%%%%%%%%%%%%%%%%%%%%%%%%%%%%%%%%%%%%%%%
\section{Data and Method}

We utilize the imaging data of HSC-SSP survey in its Wide layer, which is aiming to
observe $\sim 1,400$ deg$^2$ in five photometric bands ($g$, $r$, $i$, $z$, and $y$)
(for details, see \cite{Aihara2017a,Aihara2017b}).
The target 5$\sigma$ point-source limiting magnitudes in this layer are
($g$, $r$, $i$, $z$, $y$) = (26.5, 26.1, 25.9, 25.1, 24.4) mag.
In this paper, we use the $g$, $r$, and $i$-band data obtained
before 2016 April (internal data release S16A), covering $\sim 300$ deg$^2$ in
6 fields along the celestial equator (named XMM, WIDE12H, WIDE01H, VVDS, GAMA15H,
and GAMA09H) as well as one field around (RA,DEC)$=(242^{\circ},43^{\circ})$ (HECTOMAP).
For WIDE01H, only $g$ and $r$-band data are available in S16A.
The HSC data are processed with
hscPipe v4.0.2 \citep{Bosch2017}, a branch of the Large Synoptic Survey Telescope pipeline
\citep{Ivezic2008,Axelrod2010,Juric2015} calibrated against Pan-STARRS1
photometry and astrometry \citep{Schlafly2012,Tonry2012,Magnier2013}.
All the photometry data are corrected for the mean Galactic foreground
extinction, $A_V$ \citep{Schlafly2011}. We note that most of the dust should be
closer than the UFDs we are looking for.

\subsection{Selection of target stars}

For the purpose of searching efficiently for new faint satellites in the MW halo,
we select stars from the HSC data as follows:
(1) their images are point-like to avoid galaxies, (2) their $g-r$ colors are
bluer than 1 to eliminate foreground M-type disk stars, and (3) their $g-r$ vs. $r-i$
colors follow the fiducial relation expected for stars to remove the remaining contaminants.

First, to select point sources, we adopt the {\it extendedness} parameter from the pipeline
following our previous paper \citep{Homma2016}. In brief, this parameter is computed based on
the ratio between PSF and cmodel fluxes \citep{Abazajian2004}, where a point source has this
ratio larger than 0.985. We use this parameter measured in the $i$-band, in which the seeing
is typically the best of our five filters with a median of about $0''.6$.
As detailed in \citet{Aihara2017a},
using the combination of the HSC COSMOS and HST/ACS data \citep{Leauthaud2007},
the completeness and contamination of this star/galaxy classification is defined and quantified
as follows. The completeness, defined as the fraction of objects that are classified as stars by ACS,
and correctly classified as stars by HSC, is above 90\% at $i<22.5$, and drops to $\sim50\%$
at $i=24.5$. The contamination, defined as the fraction of HSC-classified stars
which are classified as galaxies by ACS, is close to zero at $i<23$, but increases to
$\sim 50\%$ at $i=24.5$. In this work, we adopt the extendedness parameter down to $i=24.5$
to select stars.

%%% Fig.  %%%%%%%%%%%%%%%%%%%%%%%%%%%
\begin{figure}[t!]
%\centering
\begin{center}
\includegraphics[width=80mm]{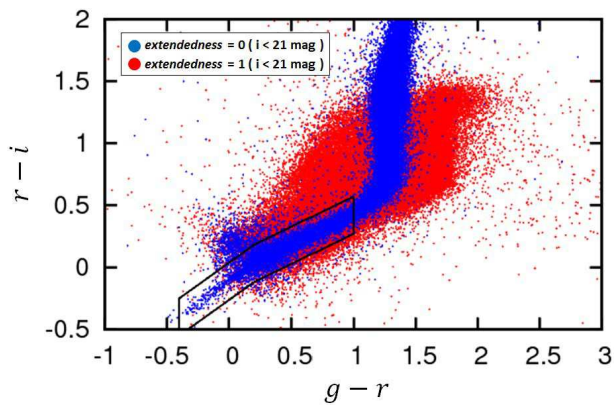}
\end{center}
\caption{
Color-color diagram of bright sources with $i < 21$~mag in WIDE12H. Blue and red dots denote
point sources ({\it extendedness}$=0$) and galaxies ({\it extendedness}$=1$), respectively.
The polygon outlines the color cut we use for the selection of target stars in the MW halo.
}
\label{fig: color-color}
\end{figure}
%%%%%%%%%%%%%%%%%%%%%%%%%%%%%%%%%%%%%%

We then use the color data to eliminate the remaining contaminants, including foreground
disk stars and background quasars and compact galaxies, which remain after the {\it extendedness} cut.
For this purpose, we plot, in Figure \ref{fig: color-color}, the $g-r$ vs.
$r-i$ relation for both stars ({\it extendedness}$=0$) and galaxies ({\it extendedness}$=1$)
in the WIDE12H field with brightness of $i < 21$~mag, where star/galaxy classification
is very reliable. As is clear, star-like objects show a narrow, characteristic sequence in this
color-color diagram compared to galaxies. Thus to optimize the selection of stars further from
the sample of point sources with {\it extendedness}$=0$, we set a color cut based on this 
color-color diagram, as has been done in previous work (e.g., \cite{Willman2002}).
Namely, we first eliminate the numerous red-color stars with
$g-r \ge 1$ in Figure \ref{fig: color-color}, which are dominated by M-type stars
in the MW disk. Then, as most likely star candidates in the MW halo, we select
point sources inside a polygon in the diagram, which is bounded by
($g-r$, $r-i$)$=$(1.00, 0.27), (1.00, 0.57), ($-0.4$, $-0.55$), and ($-0.4$, $-0.25$).
We note that the width of this color cut in $r-i$, 0.3 mag, is wider than the typical photometric
error of $r-i$ at $i = 24.5$, so as to optimally include candidate stars in the MW halo.
Although this color cut is not perfect in removing all the quasars and galaxies,
we adopt point sources selected from this color cut to search for
any signatures of overdensities as described in the next subsection.

\subsection{Detection algorithm for stellar overdensities}

Our targets here are old, metal-poor stellar systems at a particular distance,
which reveal their presence as statistically significant spatial overdensities
relative to the foreground and background noise from MW stars and distant
galaxies/quasars. For this purpose, we adopt the algorithm below, following
several other works \citep{Koposov2008,Walsh2009}.

%%% Fig.  %%%%%%%%%%%%%%%%%%%%%%%%%%%
\begin{figure}[t!]
%\centering
\begin{center}
\includegraphics[width=80mm]{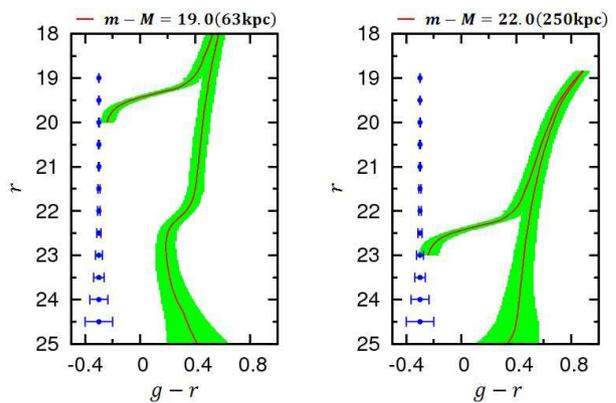}
\end{center}
\caption{
Example of a CMD filter (green shaded area) including an isochrone (red line) for an old,
metal-poor system [age of 13~Gyr and metallicity of [M$/$H]$=-2.2$]
at a distance of 63~kpc (left) and 250~kpc (right panel), respectively.
The error bars show a typical measurement error in color at each $r$ magnitude.
}
\label{fig: CMDfilter}
\end{figure}
%%%%%%%%%%%%%%%%%%%%%%%%%%%%%%%%%%%%%%

\subsubsection{Isochrone filter}

Ultra-faint dwarf satellites discovered so far show a characteristic
locus in the color-magnitude (CMD) diagram, which is similar to that of an old
MW globular cluster, namely an old, metal-poor stellar system. Thus to enhance
the clustering signal of stars in a satellite over the background noise, we select
stars which are included within a CMD locus defined for an old stellar population
within a particular distance range from the Sun. This isochrone filter is made based on
a PARSEC isochrone \citep{Bressan2012}, in which
we assume an age of 13~Gyr and metallicity of $z = 0.0001$ ([M$/$H]$=-2.2$).
We adopt a CMD defined with $g-r$ vs. $r$-band absolute magnitude, $M_r$, and
convert the SDSS filter system available for the PARSEC isochrone to
the current HSC filter system,
$g = g_{\rm SDSS} - a (g_{\rm SDSS} - r_{\rm SDSS}) - b$ and
$r = r_{\rm SDSS} - c (r_{\rm SDSS} - i_{\rm SDSS}) - d$, where
$(a,b,c,d)=(0.074,0.011,0.004,0.001)$ and the subscript SDSS denotes
the SDSS system \citep{Homma2016}. 

We then set the finite width for the selection filter as a function of $r$-band
magnitude, which is the quadrature sum of a 1$\sigma$ $(g-r)$ color measurement error
in HSC imaging and a typical color dispersion of about $\pm 0.05$ mag
at the location of the RGB arising from a metallicity dispersion of
$\pm 0.5$ dex, which is typically found in dSphs.
We shift this isochrone filter over distance moduli of $(m-M)_0 =16.5$ to 24.0
in 0.5 mag steps, which corresponds to searching for old stellar systems
with the heliocentric distance, $D_{\odot}$, between 20~kpc and 631~kpc.

Figure \ref{fig: CMDfilter} shows two examples of a CMD isochrone filter
placed at $(m-M)_0 = 19.0$ ($D_{\odot} = 63$~kpc) and $(m-M)_0 =22.0$ (250~kpc).
In the former case for a relatively nearby stellar system, it is possible
to detect the CMD feature near the main sequence turn-off, whereas
the latter more distant system shows only a red giant branch and horizontal
branch features.

\subsubsection{Search for overdensities and their statistical significance}

After selecting stars using the above isochrone filter at each distance,
we search for the signature of
their spatial overdensities and examine their statistical significance.
We count selected stars in $0^{\circ}.05 \times 0^{\circ}.05$ bins in right ascension
and declination, with an overlap of $0^{\circ}.025$ in each direction.
Here, the grid interval of $0^{\circ}.05$ corresponds to a typical half-light diameter
($\sim 80$~pc) of a ultra-faint dwarf galaxy at a distance of 90~kpc, and any signature
of dwarf galaxy at and beyond this distance, which is our target with HSC, can be
detected within this grid interval.

We count the number of stars at each cell, $n_{i,j}$, where if a cell contains
no stars, $n_{i,j}=0$, e.g. due to masking in the vicinity of a bright-star image,
we just ignore it for the following calculation. We then calculate the mean density ($\bar{n}$)
and its dispersion ($\sigma$) over all cells for each of the Wide layer fields separately
and define the normalized signal in each cell, $S_{i,j}$, giving the number of standard
deviations above the local mean (e.g., \cite{Koposov2008,Walsh2009}),
\begin{equation}
S_{i,j} = \frac{n_{i,j}-\bar{n}}{\sigma} \ .
\end{equation}
We find that the distribution of $S$ is almost Gaussian.

In order to find candidate overdensities that are statistically high enough to
reject false detections, it is necessary to set detection thresholds for the value
of $S$ \citep{Walsh2009}. For this purpose, we obtain the characteristic distribution
of $S$ for purely random fluctuations in stellar densities as follows.
First, we define an area with $\Delta {\rm RA}=10$~deg and $\Delta {\rm DEC}=5$~deg
in each of the HSC Wide-layer survey fields (corresponding to the typical survey area
of $\sim 50$~deg$^2$). In each field, stars are randomly distributed to reproduce
the mean number density, $\bar{n}$, by counting stars in each cell; again
we ignore cells with no stars $n_{i,j}=0$ in the calculation.
We then estimate the maximum values of $S$, $S_{\rm max}$, and repeat this experiment
1,000 times to calculate the mean of $S_{\rm max}$ as a function of $\bar{n}$.

\begin{figure}[t!]
%\centering
\begin{center}
\includegraphics[width=80mm]{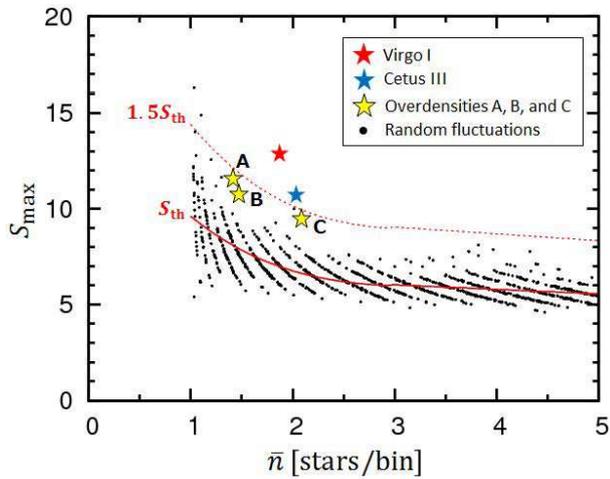}
\end{center}
\caption{
Distribution of the maximum density contrast, $S_{\rm max}$, as a function of $\bar{n}$
arising from purely random fluctuations in stellar densities (black dots).
Solid red line shows the approximation to the mean relation, $S_{\rm th}(\bar{n})$,
given as $S_{\rm th}(\bar{n}) =
-0.22 \bar{n}^3 + 2.36 \bar{n}^2 - 8.37 \bar{n} + 15.84$ for $\bar{n} < 3$ and 
$S_{\rm th}(\bar{n}) = -0.24 \bar{n} + 6.74$ for $\bar{n} \ge 3$.
Dotted red line shows $1.5 \times S_{\rm th}(\bar{n})$, which is the detection threshold 
for candidate satelites adopted in this work.
Red and blue stars denote Virgo~I and Cetus~III, respectively, whereas yellow stars
denote other three overdensities A, B, and C.
}
\label{fig: random}
\end{figure}

Figure \ref{fig: random} shows the result of this experiment.
The solid red line shows the approximation to
the mean relation, $S_{\rm th}(\bar{n})$, given as $S_{\rm th}(\bar{n}) =
-0.22 \bar{n}^3 + 2.36 \bar{n}^2 - 8.37 \bar{n} + 15.84$ for $\bar{n} < 3$ and 
$S_{\rm th}(\bar{n}) = -0.24 \bar{n} + 6.74$ for $\bar{n} \ge 3$.
We note that a typical value of $\bar{n}$ in the survey fields 
is 1 to 2, whereas $\bar{n}$ in GAMA15H is much larger, 7 at $(m-M)_0=18$ and
$\sim 4$ at $(m-M)_0 \ge 19.5$ because of the presence of the so-called
Virgo Overdensity (e.g., \cite{Juric2008}). As Figure \ref{fig: random} shows,
$S_{\rm th}$ ranges from $\sim 10$ to 7 for $\bar{n}=1$ to 2 and $S_{\rm th} \simeq 6$
for $\bar{n} \ge 4$. The dotted red line in Figure \ref{fig: random} shows
$1.5 \times S_{\rm th}$, which lies beyond basically all of the distribution
for these purely random fluctuations, except two points at $\bar{n} \simeq 1$
having high $S_{\rm th}$. Thus, in this work,
we adopt the optimal density threshold of $1.5 \times S_{\rm th}$ so as to
retain promising candidate overdensities from the currently available HSC data,
while keeping caution in interpretating the results.

We have found two candidate overdensities above
this detection threshold. The highest signal is from Virgo~I
($S=12.8$ with $\bar{n}=1.87$, $S/S_{\rm th}=1.90$) as reported in our previous paper
\citep{Homma2016}. The second highest signal is found in the direction of Cetus
($S=10.7$ with $\bar{n}=2.02$, $S/S_{\rm th}=1.58$).
There are other three overdensities with relatively high signal of
$S/S_{\rm th}=1.3 \sim 1.4$, hereafter denoted as overdensities A, B, and C,
which however may be false detections as judged from several other properties
as explained below.

%%% Fig.  %%%%%%%%%%%%%%%%%%%%%%%%%%%
%%%%%%%%%%%%%%%%%%%%%%%%%%%%%%%%%%%%%%

%%% Sec.3 %%%%%%%%%%%%%%%%%%%%%%%%%%%%%%%%%%%%%%%%%%%%%%%%
\section{Results}

Following the procedure described in the previous section, we have found
a highly compelling dwarf galaxy candidate in the direction of Cetus,
hereafter Cetus~III,
in addition to Virgo~I, and also identified other three overdensities that
appear to be false detections. We have also detected known substructures such
as known globular clusters, which have a high density signal, which are removed
from the following analysis. In this section, we confine ourselves to describe
Cetus~III in detail and briefly comment on other overdensities showing
high signal.

\subsection{HSC~$J0209-0416$ - a new satellite candidate, Cetus~III}

\begin{figure*}[t!]
%\centering
\begin{center}
\includegraphics[width=120mm]{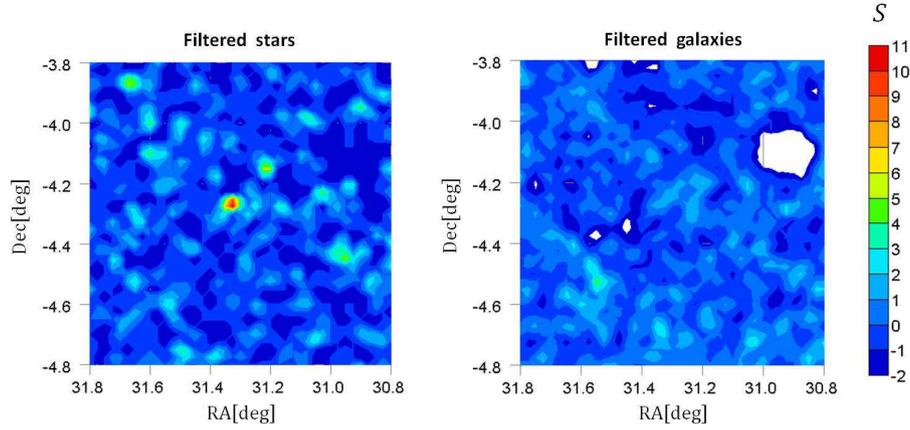}
\end{center}
\caption{
Left panel: the spatial distribution of the stellar overdensity in Cetus
passing the isochrone filter at $(m-M)_0 = 22.0$ with constraints of $i < 24.5$ and $g-r<1.0$,
covering one square degree centered on this candidate.
Right panel: the plot for the sources classified as galaxies
passing the same isochrone filter and same constraints as for the stars in the left panel.
Note that there is no overdensity at the center of this plot.
}
\label{fig: Cetus_space}
\end{figure*}

This overdensity signal with $S/S_{\rm th}=1.58$ and $\bar{n}=2.02$ (10.7$\sigma$)
is found at $(\alpha, \delta)=(31^{\circ}.325, -4^{\circ}.275)$
and $(m-M)_0 = 22.0$~mag in the XMM field.
Figure \ref{fig: Cetus_space} shows this feature, which passes the isochrone filter
at the above distance modulus, for stars (left) and galaxies (right). It is clear
that there is no corresponding overdensity in galaxies.

%%% Fig.  %%%%%%%%%%%%%%%%%%%%%%%%%%%
%%%%%%%%%%%%%%%%%%%%%%%%%%%%%%%%%%%%%%

%%% Fig.  %%%%%%%%%%%%%%%%%%%%%%%%%%%
\begin{figure*}[t!]
%\centering
\begin{center}
\includegraphics[width=150mm]{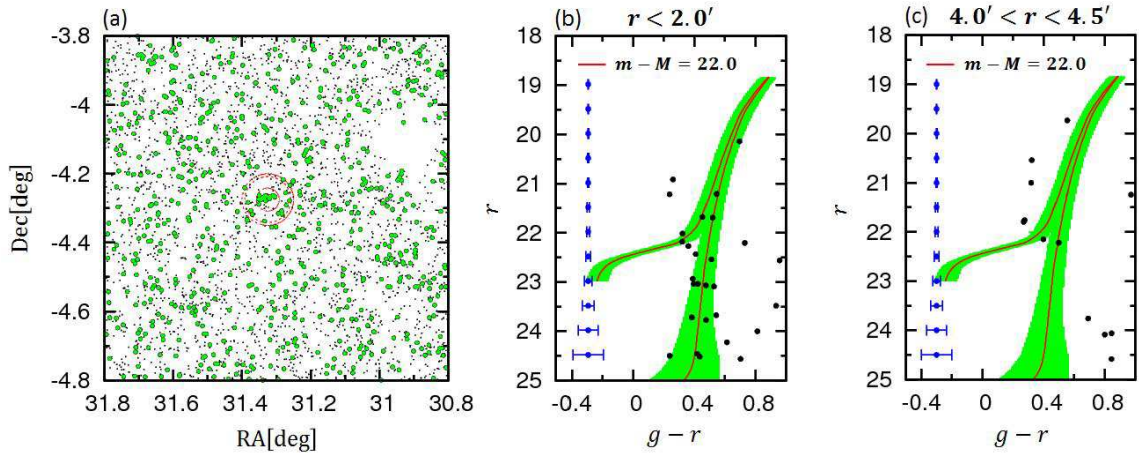}
\end{center}
\caption{
Panel (a): the spatial distribution of the stars around the overdensity in Cetus, where
green circles (black dots) denote the stars inside (outside) the isochrone filter
at $(m-M)_0 = 22.0$.
Red circles denote annuli with radii $r=2'$, $4'$, and $4'.5$ from the center.
Panel (b): the distribution of the stars in the $g-r$ vs. $r$ CMD at $r < 2'$.
Panel (c): the same as (b) but for field stars at $4' < r < 4'.5$,
which has the same solid angle. Note the absence of a red giant branch.
}
\label{fig: Cetus_cmd}
\end{figure*}
%%%%%%%%%%%%%%%%%%%%%%%%%%%%%%%%%%%%%%

In Figure \ref{fig: Cetus_cmd}(a), we plot the spatial distribution of all the stars
around this overdensity, which shows a localized concentration of stars within a circle of
radius $2'$. Panel (b) shows the $(g-r, r)$ CMD of stars within the $2'$ radius circle
shown in panel (a). This CMD shows a clear signature of a red giant branch (RGB), whereas
this feature disappears when we plot stars at $4' < r < 4'.5$ with the same solid angle,
i.e. likely field stars outside the overdensity, as shown in panel (c).

\subsubsection{Distance estimate}

\begin{figure}[t!]
%\centering
\begin{center}
\includegraphics[width=80mm]{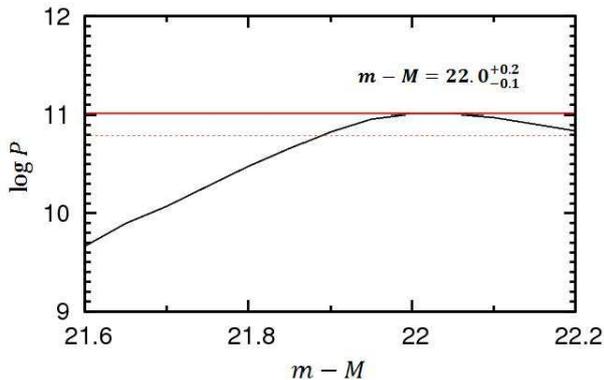}
\end{center}
\caption{
Probability distribution of the 15 likely member stars at $r < 2'$ relative to the best-fit
isochrone case at $(m-M)_0 =22.0$ as a function of $(m-M)_0$. The maximum probability is $\log P = 11.01$
delineated with red solid line and 1$\sigma$ interval is denoted by red dotted line.
}
\label{fig: Cetus_distance}
\end{figure}

The heliocentric distance to this stellar system is derived based on the likelihood analysis,
for which we use 15 likely member RGBs inside the isochrone envelope in Figure \ref{fig: Cetus_cmd}(b)
inside $r < 2'$ at the best-fit case of $(m-M)_0 = 22.0$ in the search of the overdensity.
Assuming that the probability distribution, $P$, of these member stars relative to the best-fit
isochrone in the CMD is Gaussian, we obtain the dependence of $P$ on $(m-M)_0$ as shown in
Figure \ref{fig: Cetus_distance}. We thus arrive at the distance modulus of Cetus~III of
$(m-M)_0 = 22.0^{+0.2}_{-0.1}$, corresponding to the heliocentric distance of $D = 251^{+24}_{-11}$~kpc.

%%% Fig. 3 %%%%%%%%%%%%%%%%%%%%%%%%%%%
%%%%%%%%%%%%%%%%%%%%%%%%%%%%%%%%%%%%%%

\subsubsection{Structural parameters}

We estimate the structural properties of this overdensity following \citet{Martin2008,
Martin2016}. We set six parameters $(\alpha_0, \delta_0, \theta, \epsilon,
r_h, N_{\ast})$: $(\alpha_0, \delta_0)$ for the celestial coordinates
of the centroid of the overdensity, $\theta$ for its position angle from north to
east, $\epsilon$ for the ellipticity, $r_h$ for the half-light radius
measured on the major axis, and
$N_{\ast}$ for the number of stars belonging to the overdensity. 
The maximum likelihood method of \citet{Martin2008} is applied to the stars
within a circle of radius $8'$ (corresponding to about 5.6 times of the anticipated
$r_h$) passing the isochrone filter; the results are summarized in Table~\ref{tab: 1}.
It is worth noting that this stellar system is characterized by its small number of
stars with $N_{\ast}=16^{+3}_{-5}$ and ellipticity of $\epsilon =0.76$.

%%% Tab 1%%%%%%%%%%%%%%%%%%%%%%%%%%%%%
\begin{table}
\tbl{Properties of Cetus~III}{
%\tablewidth{0pt}
\begin{tabular}{lc}
\hline
Parameter$^{a}$ & Value \\
\hline
Coordinates (J2000)           & $31^{\circ}.331$, $-4^{\circ}.270$        \\
Galactic Coordinates ($l,b$)  & 163$^{\circ}$.810, $-61^{\circ}.133$      \\
Position angle                & $+101^{+5}_{-6}$ deg                      \\
Ellipticity                   & $0.76^{+0.06}_{-0.08}$                    \\
Number of stars, $N_{\ast}$   & $16^{+3}_{-5}$                            \\
$A_V$                         & 0.066 mag                                 \\
$(m-M)_0$                     & 22.0$^{+0.2}_{-0.1}$ mag                  \\
Heliocentric distance         & 251$^{+24}_{-11}$ kpc                             \\
Half light radius, $r_h$      & $1'.23^{+0'.42}_{-0'.19}$ or 90$^{+42}_{-17}$ pc    \\
$M_{{\rm tot},V}$             & $-2.45^{+0.57}_{-0.56}$ mag  \\
\hline
\end{tabular}}\label{tab: 1}
\begin{tabnote}
$^{a}$Integrated magnitudes are corrected for the mean Galactic foreground extinction,
$A_V$ \citep{Schlafly2011}.
\end{tabnote}
\end{table}
%%%%%%%%%%%%%%%%%%%%%%%%%%%%%%%%%%%%%%

%%% Fig.  %%%%%%%%%%%%%%%%%%%%%%%%%%%
\begin{figure}[t!]
%\centering
\begin{center}
\includegraphics[width=80mm]{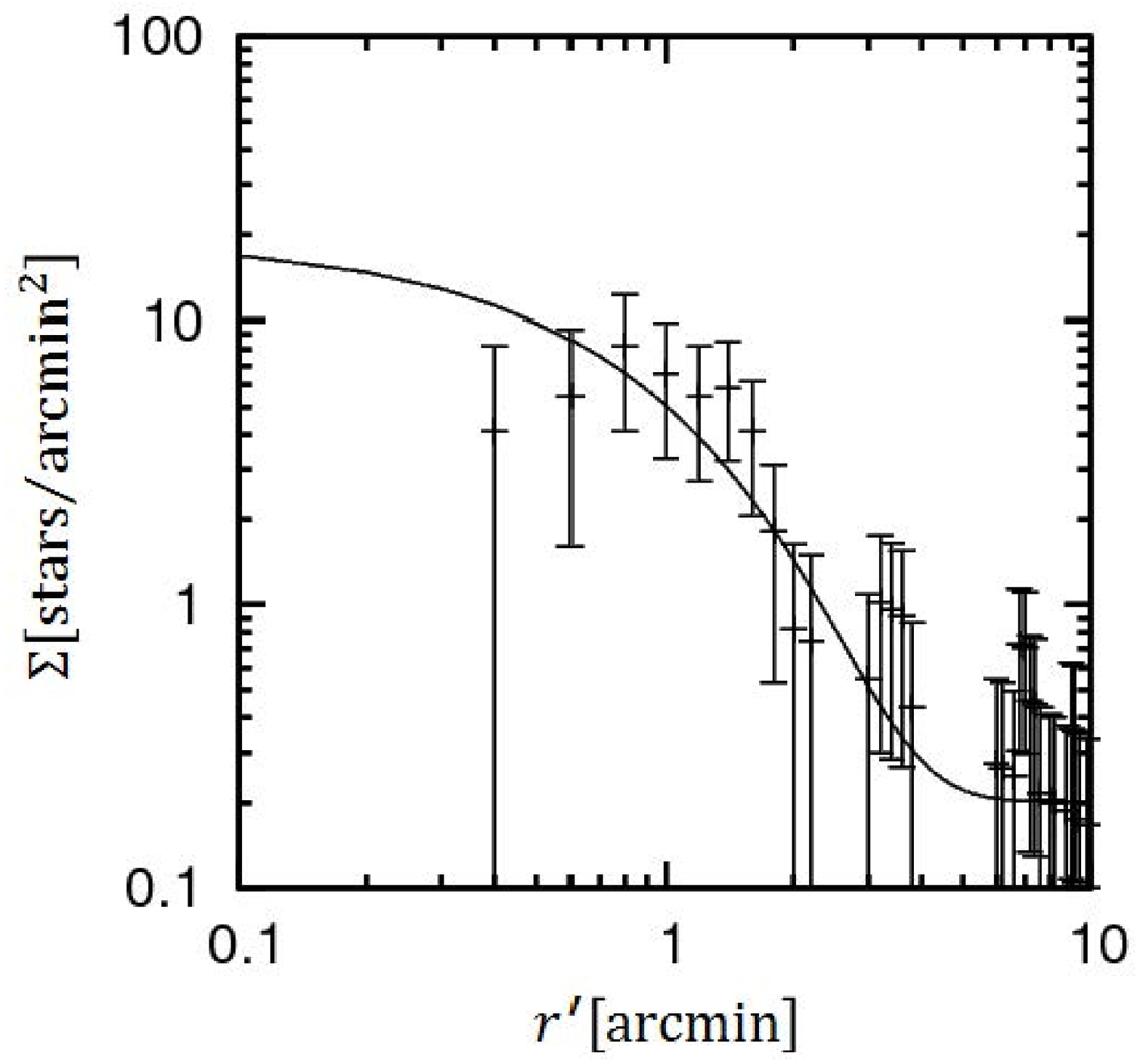}
\end{center}
\caption{Density profile of the stars in Cetus~III that pass the isochrone filter
shown in Figure \ref{fig: Cetus_cmd}(b), in elliptical annuli as a function of mean radius,
where the uncertainties are derived assuming Poisson statistics.
The line shows a fitted exponential profile with $r_h = 1'.23$.}
\label{fig: Cetus_profile}
\end{figure}
%%%%%%%%%%%%%%%%%%%%%%%%%%%%%%%%%%%%%%

Figure \ref{fig: Cetus_profile} shows the radial profile of the stars passing the isochrone
filter [Figure \ref{fig: Cetus_cmd}(b)] by computing the average density within
elliptical annuli. The overplotted line corresponds to the best-fit
exponential profile with a half-light radius of $r_h = 1'.23$ or 90~pc.
This spatial size is larger than the typical size of MW globular clusters
but is consistent with the scale of dwarf satellites as examined below.

\subsubsection{$V$-band absolute magnitude}

The $V$-band absolute magnitude of Ceus~III, $M_V$, is estimated in several
ways as follows, where for the transformation from $(g, r)$ to $V$, we adopt
the formula in \citet{Jordi2006} calibrated for metal-poor Population II stars.

The simplest method is just to sum the luminosities of the stars
within the half-light radius, $r_h$, and then doubling the summed luminosity
(e.g., \cite{Sakamoto2006}). Using the best-fit distance of $(m-M)_0 = 22.0$~mag,
we obtain $M_V = -3.04^{+0.26}_{-0.39}$ mag for $r_h = 1'.23^{+0'.42}_{-0'.19}$
and $(m-M)_0 = 22.0^{+0.2}_{-0.1}$.

However this method suffers from shot noise due to the small number of stars
in Cetus~III, which is a significant additional source of uncertainty in estimating
$M_V$. Thus, as was done in our previous paper \citep{Homma2016}, we adopt
a Monte Carlo method similar to that described in \citet{Martin2008} to determine
the most likely value of $M_V$ and its uncertainty. 
Based on the values of $N_{\ast} = 16^{+3}_{-5}$ at $i < 24.5$ mag and
$(m-M)_0 = 22.0^{+0.2}_{-0.1}$ mag obtained in the previous subsection
and on a stellar population model with an age of 13~Gyr and
metallicity of [M$/$H]$=-2.2$, we generate $10^4$ realizations of CMDs for
the initial mass function (IMF) by \citet{Kroupa2002}.
We then derive the luminosity of the stars at $i < 24.5$ mag, taking into
account the completeness of the observed stars with HSC.
The median (mean) value of $M_V$ is then given as
$M_V = -2.45^{+0.57}_{-0.56}$~mag ($M_V = -2.58^{+0.69}_{-0.34}$~mag).
When we consider unobserved faint stars below the limit of $i=24.5$,
the $M_V$ estimate becomes slightly brighter,
$M_V = -3.09^{+0.57}_{-0.44}$~mag ($M_V = -3.14^{+0.49}_{-0.34}$~mag).
Note that all of these $M_V$ values are consistent within the 1$\sigma$
uncertainty and also are insensitive to the adoption of different
IMFs such as Salpeter and Chabrier IMFs \citep{Salpeter1955,Chabrier2001}.

In deriving $M_V$ corrected for unobserved faint stars,
we are also able to obtain the total mass of stars in Cetus~III, $M_{\ast}$,
for different IMFs. They are summarized as
$M_{\ast}=2300^{+917}_{-785} M_{\odot}$ (Kroupa),
$4302^{+1833}_{-1721} M_{\odot}$ (Salpeter), and $2205^{+894}_{-771} M_{\odot}$ (Chabrier).

%%% Tab 2%%%%%%%%%%%%%%%%%%%%%%%%%%%%%
\begin{table}
\tbl{Revised Properties of Virgo~I}{
%\tablewidth{0pt}
\begin{tabular}{lc}
\hline
Parameter$^{a}$ & Value \\
\hline
Coordinates (J2000)           & $180^{\circ}.038$, $-0^{\circ}.681$       \\
Galactic Coordinates ($l,b$)  & 276$^{\circ}$.942, $+59^{\circ}.578$      \\
Position angle                & $+62^{+8}_{-13}$ deg                      \\
Ellipticity                   & $0.59^{+0.12}_{-0.14}$                    \\
Number of stars, $N_{\ast}$   & $18^{+5}_{-4}$                            \\
$A_V$                         & 0.066 mag                                 \\
$(m-M)_0$                     & 19.8$^{+0.2}_{-0.1}$ mag                  \\
Heliocentric distance         & 91$^{+9}_{-4}$ kpc                             \\
Half light radius, $r_h$      & $1'.76^{+0'.49}_{-0'.40}$ or 47$^{+19}_{-13}$ pc    \\
$M_{{\rm tot},V}$             & $-0.33^{+0.75}_{-0.87}$ mag \\
\hline
\end{tabular}}\label{tab: 2}
\begin{tabnote}
$^{a}$Integrated magnitudes are corrected for the mean Galactic foreground extinction,
$A_V$ \citep{Schlafly2011}.
\end{tabnote}
\end{table}
%%%%%%%%%%%%%%%%%%%%%%%%%%%%%%%%%%%%%%

\subsection{Properties of other overdensities showing high signal}

\subsubsection{HSC~$J1200-0040$ - Virgo~I}

This new satellite already reported in \citet{Homma2016} shows the highest signal of
$S= 12.8$ with $\bar{n}=1.87$ and $S/S_{\rm th}=1.90$.
In contrast to our previous paper \citep{Homma2016}, we have adopted the $i$-band data
for the removal of contaminations in addition to the $g$ and $r$-band data, so that
the final results for the physical parameters of stellar system are slightly changed.
We thus list the revised values for Virgo~I in Table~\ref{tab: 2}, although the differences
are well within the 1$\sigma$ uncertainty. For the median (mean) value of $M_V$
with completeness correction, we obtain
$M_V = -0.33^{+0.75}_{-0.87}$~mag ($M_V = -0.76^{+1.18}_{-0.45}$~mag)
\footnote{In \citet{Homma2016}, we reported only the mean $M_V$ of $\sim -0.8$ mag.}.
When we correct for unobserved faint stars below the limit of $i=24.5$, we obtain
$M_V = -0.70^{+0.55}_{-0.69}$~mag ($M_V = -1.04^{+0.80}_{-0.38}$~mag).
The total stellar mass of Virgo~I is estimated as
$M_{\ast}=323^{+129}_{-108} M_{\odot}$ (Kroupa),
$625^{+255}_{-211} M_{\odot}$ (Salpeter), and $313^{+128}_{-101} M_{\odot}$ (Chabrier).

\subsubsection{Other three high signals with possibly false detections}

\begin{figure*}[t!]
%\centering
\begin{center}
\includegraphics[width=160mm]{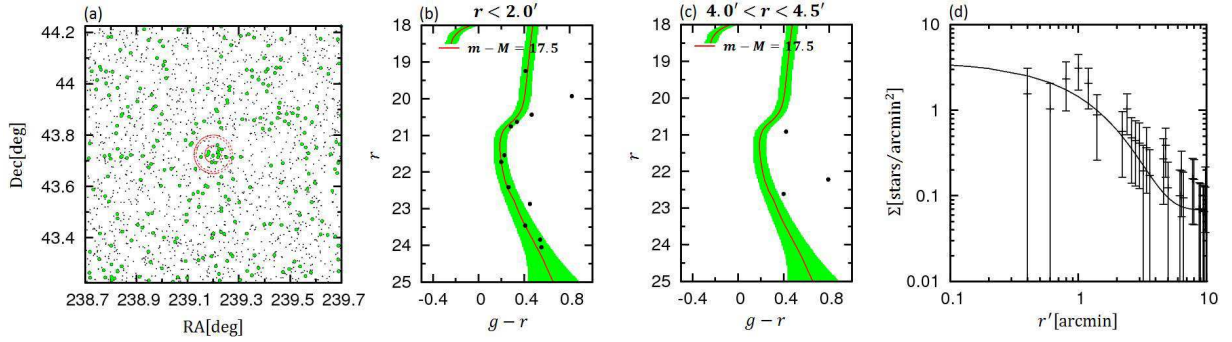}
\end{center}
\caption{
Panel (a): the spatial distribution of stars in the overdensity A with $S/S_{\rm th}=1.39$ found at 
(RA,DEC)$=(239^{\circ}.200, 43^{\circ}.725)$, where
red circles denote annuli with radii $r=2'$, $4'$, and $4'.5$ from the center.
Panel (b): the distribution of the stars in the $g-r$ vs. $r$ CMD at $r < 2'$.
Panel (c): the same as (b) but for field stars at $4' < r < 4'.5$.
Panel (d): the density profile of the stars that pass the isochrone filter in (b),
in elliptical annuli as a function of mean radius.
where the uncertainties are derived assuming Poisson statistics.
The line shows a fitted exponential profile.
}
\label{fig: fake1}
\end{figure*}

\begin{figure*}[t!]
%\centering
\begin{center}
\includegraphics[width=160mm]{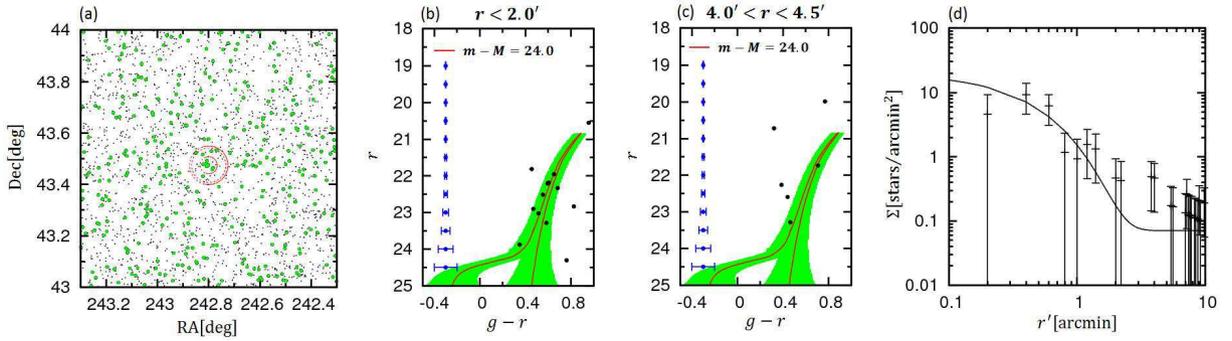}
\end{center}
\caption{
The same as Figure \ref{fig: fake1} but for the overdensity B with $S/S_{\rm th}=1.33$
found at (RA,DEC)$=(242^{\circ}.800, 43^{\circ}.475)$.
}
\label{fig: fake2}
\end{figure*}

\begin{figure*}[t!]
%\centering
\begin{center}
\includegraphics[width=160mm]{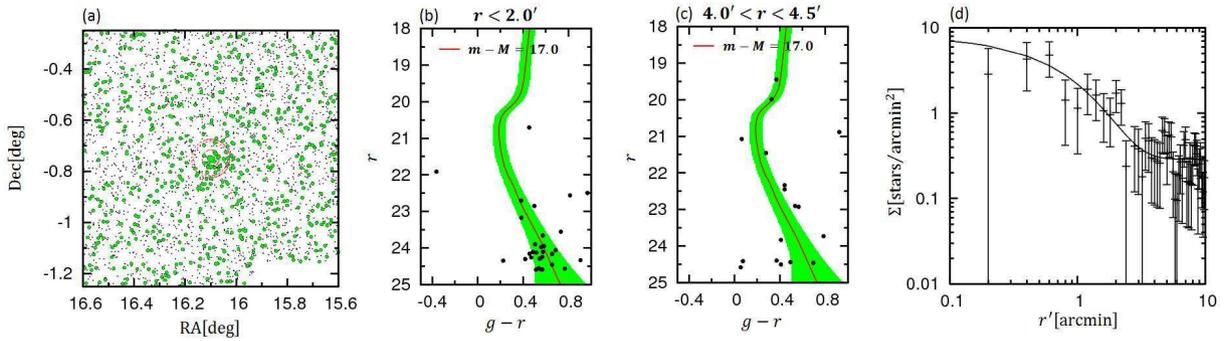}
\end{center}
\caption{
The same as Figure \ref{fig: fake1} but for the overdensity C with $S/S_{\rm th}=1.41$
found at (RA,DEC)$=(16^{\circ}.100, -0^{\circ}.750)$.
}
\label{fig: fake3}
\end{figure*}

We summarize the status of the other three overdensities, A, B, and C,  found from
their relatively high signals of $S/S_{\rm th}$. They are plotted in Figure \ref{fig: fake1},
\ref{fig: fake2} and \ref{fig: fake3}.

\begin{itemize}
\item A: (RA,DEC)$=(239^{\circ}.200, 43^{\circ}.725)$ (in HECTOMAP) at $(m-M)_0=17.5$ giving
$S/S_{\rm th}=1.39$ with $\bar{n}=1.40$ (Figure \ref{fig: fake1}): the suface number
density of stars is even much smaller than Cetus~III. The paucity of main-sequence
stars even at $r<2'$ and the deviation from an exponential profile, with a spiky feature
near $1'$ from the center, makes it difficult to conclude that this overdensity is
indeed a stellar system.
\item B: (RA,DEC)$=(242^{\circ}.800, 43^{\circ}.475)$ (in HECTOMAP) at $(m-M)_0=24.0$ giving
$S/S_{\rm th}=1.33$ with $\bar{n}=1.44$ (Figure \ref{fig: fake2}): this somewhat high signal is
given only by the effect of a few stars in the bright RGB locus at $r<2'$.
Their spatial distribution shows somewhat extended outskirts and thus deviates
from an exponential profile.
It is thus difficult to conclude robustly that this feature is the signature of a stellar system.
Deeper imaging with space telescopes would be worthwhile for trying to identify fainter
RGB stars against background faint galaxies.
\item C: (RA,DEC)$=(16^{\circ}.100, -0^{\circ}.750)$ (in WIDE01H) at $(m-M)_0=17.0$ giving
$S/S_{\rm th}=1.41$ with $\bar{n}=2.07$ (Figure \ref{fig: fake3}): since this field still lacks
data in $i$ band, the removal of background galaxies based on
the color cut is incomplete. The CMD at $r<2'$ also implies that the signal is due to the
contamination of galaxies.
\end{itemize}

%%% Fig.  %%%%%%%%%%%%%%%%%%%%%%%%%%%
%%%%%%%%%%%%%%%%%%%%%%%%%%%%%%%%%%%%%%
%%% Fig.  %%%%%%%%%%%%%%%%%%%%%%%%%%%
%%%%%%%%%%%%%%%%%%%%%%%%%%%%%%%%%%%%%%
%%% Fig.  %%%%%%%%%%%%%%%%%%%%%%%%%%%
%%%%%%%%%%%%%%%%%%%%%%%%%%%%%%%%%%%%%%

%%% Sec.4 %%%%%%%%%%%%%%%%%%%%%%%%%%%%%%%%%%%%%%%%%%%%%%%%
\section{Discussion}

\subsection{Comparison with globular clusters and other satellites}

\begin{figure}[t!]
%\centering
\begin{center}
\includegraphics[width=80mm]{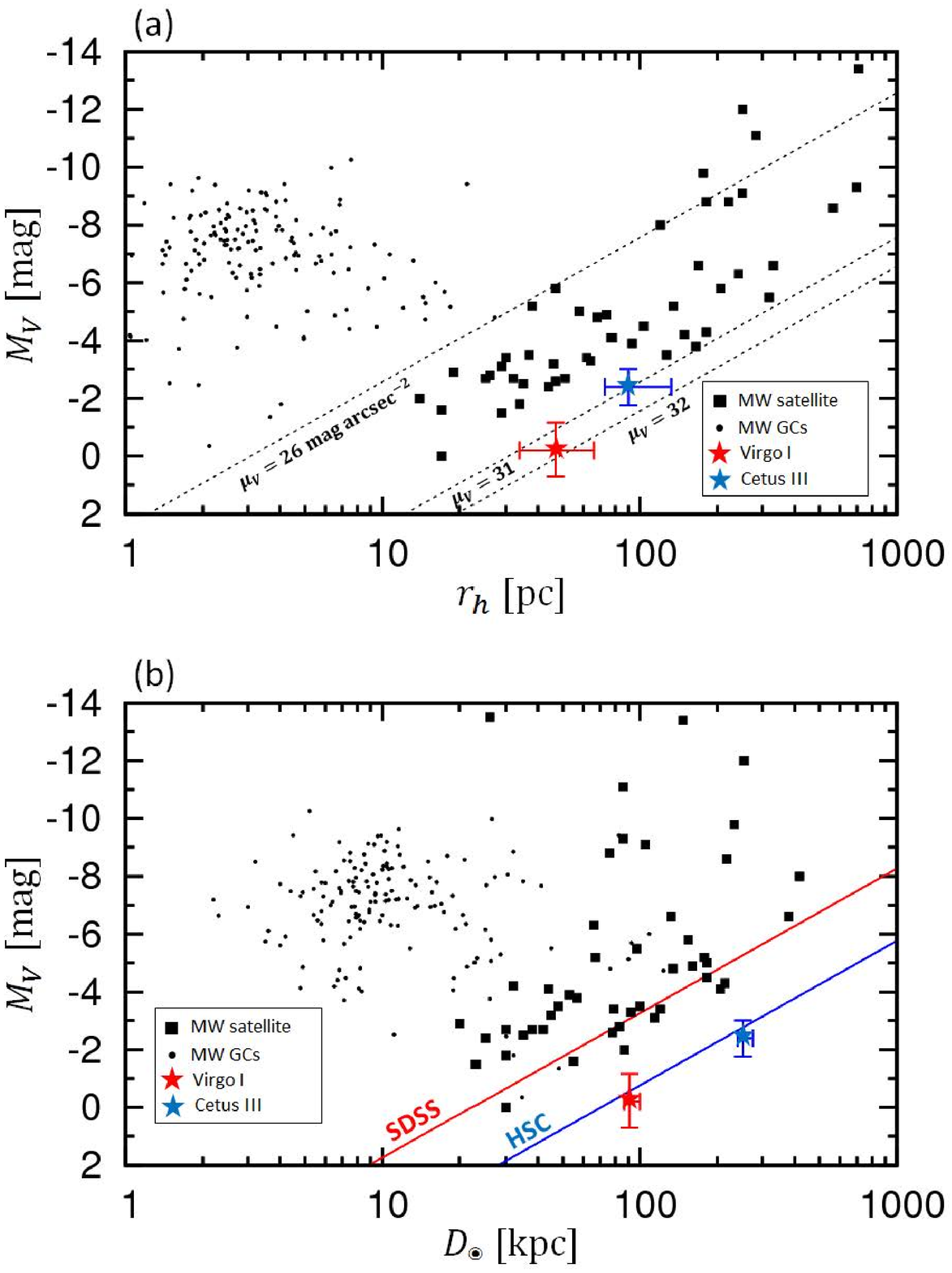}
\end{center}
\caption{(a) The relation between $M_V$ and $r_h$ for stellar systems. Dots denote
globular clusters in the MW taken from \citet{Harris1996}. Filled squares
denote the MW dSphs taken from \citet{McConnachie2012},
the recent DES work for new ultra-faint MW dSphs
\citep{Bechtol2015,Koposov2015,Drlica-Wagner2015},
and other recent discoveries \citep{Laevens2014,Kim2015,KimJerjen2015,
Laevens2015a,Laevens2015b}.
The red and blue stars with error bars correspond to Virgo~I and Cetus~III,
respectively, both of which lie within the locus defined by dSphs.
The dotted lines indicate the constant surface brightness, $\mu_V = 26$, 31,
and 32 mag~arcsec$^{-2}$.
(b) The relation between $M_V$ and heliocentric distance for the systems shown
in panel (a). The red and blue lines indicate the detection limits of SDSS and HSC,
respectively.
}
\label{fig: abs_size}
\end{figure}

Figure \ref{fig: abs_size}(a) shows the comparison in the size and absolute magnitude
of the stellar system, measured by $r_h$, between MW globular clusters (dots) \citep{Harris1996} and
the known dwarf satellites (squares) \citep{McConnachie2012,Bechtol2015,
Koposov2015,Drlica-Wagner2015,Laevens2014,Kim2015,KimJerjen2015,Laevens2015a,Laevens2015b}
including recent discoveries of the DES and PS1 surveys. Red and blue squares with
error bars show Virgo~I and Cetus~III, respectively.

Both Cetus~III and Virgo~I are significantly larger than
MW globular clusters with comparable $M_V$. They both are located along
the locus of MW dwarf satellites, thereby suggesting that Cetus~III is a highly compelling
ultra-faint dwarf galaxy candidate.
This stellar system is significantly flattened with an ellipticity of
$\epsilon=0.76^{+0.06}_{-0.08}$ like UMa~I with $\epsilon=0.80$ \citep{Martin2008},
which also supports this conclusion.

Figure \ref{fig: abs_size}(b) shows the relation between $M_V$ and the heliocentric
distance for MW globular clusters and dwarf satellites including Virgo~I and
Cetus~III. Red and blue lines denote the detection limits of SDSS and HSC, respectively.
As is clear, both Cetus~III at 251$^{+24}_{-11}$~kpc and Virgo~I at 91$^{+9}_{-4}$~kpc
are well beyond the detection limit of SDSS and are close to the limit of HSC.

%%% Fig.  %%%%%%%%%%%%%%%%%%%%%%%%%%%
%%%%%%%%%%%%%%%%%%%%%%%%%%%%%%%%%%%%%%

\subsection{Implication for the missing satellites problem}

We have identified two new dwarf satellites from $\sim 300$~deg$^2$ of HSC data
in the outer part of the MW halo, where no other survey programs can reach.
This suggests that we can expect to find more dwarf satellites
within the detection limit of HSC as the survey footprint grows.
In order to assess if this is consistent with the prediction of $\Lambda$CDM
models in the context of the missing
satellites problem, and determine how many more satellites will be found in
the completed HSC survey over $\sim 1400$~deg$^2$,
we examine the recent results of high-resolution N-body simulation for
$\Lambda$CDM models and estimate the most likely number of visible satellites
with HSC based on several assumptions for the baryon fraction relative to dark
matter and the spatial distribution of dark halos.

First, to derive the mass function of subhalos associated with a MW-sized host
halo, we adopt the result of the Caterpillar simulation \citep{Griffen2016} for
the cosmological evolution of cold dark matter halos \citep{Dooley2016}.
The Caterpillar simulation is suitable for our current purpose because of
its very high mass resolution of $\sim 10^4 M_{\odot}/{\rm particle}$
designed to study the properties of subhalos.
We adopt the analysis of this simulation by \citet{Dooley2016} in what follows.
The mass function of subhalos with mass $M_{\rm sub}$ associated with
a MW-sized host halo with mass $M_{\rm host}$ is given as
\begin{equation}
\frac{d\bar{N}}{dM_{\rm sub}} 
= K_0 \left( \frac{M_{\rm sub}}{M_{\odot}} \right)^{-\alpha}
  \frac{M_{\rm host}}{M_{\odot}} \ ,
\label{eq: massfun}
\end{equation}
where we adopt $K_0 = 0.00188$, $\alpha=1.87$, and
$M_{\rm host}=1.4\times 10^{12}M_{\odot}$ in what follows.

Here, the number of simulated subhalos varies with the distance, $r$, from the
center of a host halo. The cumulative number of subhalos inside $r$ in the
Caterpillar simulation is obtained from Equation (\ref{eq: massfun}) by
multiplying by the fraction of subhalos within a radius $r$, as $f_{\rm sub}(<r)$, parameterized
$f_{\rm sub}(<r) = k_1 + k_2(r/R_{\rm vir}) + k_3(r/R_{\rm vir})^2 + k_4(r/R_{\rm vir})^3$
with $(k_1,k_2,k_3,k_4)=(-0.0440,0.3913,0.9965,-0.3438)$ \citep{Dooley2016}.
We ignore the effect of any anisotropic distribution of subhalos; no anisotropy
is seen in the Caterpillar simulation.

\begin{figure}[t!]
%\centering
\begin{center}
\includegraphics[width=80mm]{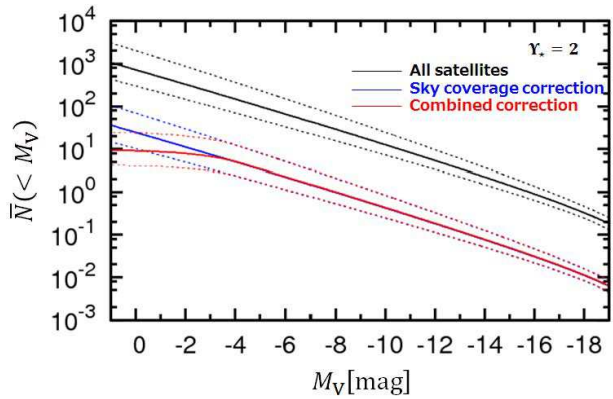}
\end{center}
\caption{
Black solide line denotes the cumulative luminosity function of visible satellites
calculated from subhalos in a MW-sized halo taken from Caterpillar simulation,
where we assume the mass to light ratio of $\Upsilon_{\ast}=2$.
Blue solid line considers the correction only for the sky coverage,
$f_{\Omega,{\rm HSC}}$, of the HSC-SSP survey, whereas red solid line considers
all the corrections associated with the HSC-SSP survey as given in
Equation (\ref{eq: corr_lumifun}). Dotted lines above and below all of black,
green and red solid lines correspond to typical errors associated with the
abundance matching method for transforming dark-halo mass to stellar mass.
}
\label{fig: lumifun}
\end{figure}

Second, based on this mass function of subhalos, we estimate the likely stellar
mass function using the abundance matching method for assigning stellar mass $M_{\ast}$
to a dark halo with mass $M_{\rm halo}$. We follow the paper by \citet{Garrison-Kimmel2017},
in which the $M_{\ast}$ vs. $M_{\rm halo}$ relation derived from the massive regime
$M_{\rm halo} > 10^{11.5}M_{\odot}$ ($M_{\ast} > 5\times 10^9 M_{\odot}$)
is extrapolated to the lower masses with finite scatter\footnote{In \citet{Garrison-Kimmel2017},
$M_{\rm halo}$ is defined as the largest instantaneous virial mass associated with the
main branch of each subhalo's merger tree, and we follow this definition in our analysis.}.
We adopt a larger scatter with decreasing $M_{\rm halo}$ in this lower mass range,
parameterized by $v=-0.2$ in their formulation [See equation (3) of \citet{Garrison-Kimmel2017}].
With this, the corresponding luminosity function of satellites, $d\bar{N}/dM_V$, can be inferred
once the mass to light ratio, $\Upsilon_{\ast}$, is given. The cumulative luminosity function
of subhalos derived in this way is shown in Figure \ref{fig: lumifun} (black solid line),
where we assume $\Upsilon_{\ast}=2$ in solar units.
This plot suggests the presence of $\sim 1000$ visible
satellites with $M_V < 0.0$ in a MW-sized halo.

Given the theoretically predicted number of visible satellites from $\Lambda$CDM models,
we consider the specific settings of HSC-SSP in its survey area and depth to estimate
the actual number of observable satellites, following the work by \citet{Tollerud2008}
estimated for SDSS. 

First, HSC-SSP in its Wide layer will observe a fraction of the sky
$f_{\Omega,{\rm HSC}} = (1400~{\rm deg}^2/41252~{\rm deg}^2) = 0.034$.
Next, we consider the completeness correction due to the detection limit of HSC,
$f_{r,{\rm HSC}}$, which thus depends on $M_V$. To do so, we use a spherical
completeness radius, $R_{\rm comp}^{\rm SDSS}(M_V)$, derived for SDSS, beyond which a satellite of
a particular magnitude is undetected \citep{Tollerud2008},
\begin{equation}
R_{\rm comp}^{\rm SDSS} (M_V) = \left( \frac{3}{4\pi f_{\rm DR5}} \right)^{1/3}
                                10^{(-aM_r-b)/3} \ {\rm Mpc} \ ,
\label{eq: Rcomp}
\end{equation}
where $f_{\rm DR5}=0.194$ is the sky coverage of SDSS DR5 and $(a,b)$ depends on
the assumed value of a virial radius.
The corresponding radius for HSC, $R_{\rm comp}^{\rm HSC}$, is given as
\begin{equation}
R_{\rm comp}^{\rm HSC}/R_{\rm comp}^{\rm SDSS} (M_V) =
                      10^{(M_{r,{\rm HSC}} - M_{r,{\rm SDSS}})/5}  \ ,
\end{equation}
where $M_{r,{\rm HSC}}$ and $M_{r,{\rm SDSS}}$ are cut-off $r$-band magnitudes
for HSC and SDSS, respectively. We set $M_{r,{\rm SDSS}}=22.2$~mag with $\sim 100$\%
completeness and $M_{r,{\rm HSC}}=24.7$~mag with $\sim 50$\% completeness,
since the cut-off magnitude in $i$ in this work is
$i = 24.5$~mag and typical satellite stars show $r-i=0.2$.
We thus obtain
\begin{equation}
f_{r,{\rm HSC}} (M_V) = f_{\rm sub} (< R_{\rm comp}^{\rm SDSS} (M_V)) \ ,
\end{equation}
where we set $R_{vir}=417$~kpc so as to include Leo~T and set $a$ and $b$
in equation (\ref{eq: Rcomp}) to $(0.684,5.667)$ \citep{Tollerud2008}.
We then derive the cumulative luminosity function of satellites
observable by HSC-SSP, as
\begin{equation}
\frac{d\bar{N}_{\rm HSC}}{dM_V} = f_{\Omega,{\rm HSC}} f_{r,{\rm HSC}} (M_V) 
       \frac{d\bar{N}}{dM_V} \ .
\label{eq: corr_lumifun}
\end{equation}

In Figure \ref{fig: lumifun}, the blue solid line corresponds to the luminosity
function corrected only for the sky coverage, $f_{\Omega,{\rm HSC}}$, and
the red solid line considers the full correction given in
equation (\ref{eq: corr_lumifun}), i.e., the expected number of satellites in the 
HSC-SSP survey. We assume here $\Upsilon_{\ast}=2$, but even if we adopt
$\Upsilon_{\ast}=1$ or 4, the number of satellites is within the uncertainty
of the abundance matching method shown with dotted lines in Figure \ref{fig: lumifun}.

Although there remain uncertainties in the prediction of visible satellites,
we anticipate that the completed 1,400~deg$^2$ HSC-SSP survey
will find $9^{+15}_{-5}$ satellites with $M_V < 0.0$.
Thus far, we have identified two new satellites over $\sim 300$~deg$^2$,
which is consistent with this prediction.
Thus, further search for new satellites from HSC-SSP survey will provide important
insights into the missing satellites problem.

%%% Fig.  %%%%%%%%%%%%%%%%%%%%%%%%%%%
%%%%%%%%%%%%%%%%%%%%%%%%%%%%%%%%%%%%%%

%%% Sec.5 %%%%%%%%%%%%%%%%%%%%%%%%%%%%%%%%%%%%%%%%%%%%%%%%
\section{Conclusions}

In this paper, we have presented our updated method, compared to our previous work
\citep{Homma2016}, for the systematic search for old stellar systems from 300~deg$^2$ of
HSC-SSP survey data and have identified a highly compelling ultra-faint dwarf
satellite candidate, Cetus~III, as a statistically high overdensity, in addition to
Virgo~I. This stellar system is a 10.7$\sigma$ overdensity in the relevant
isochrone filter in the relevant survey area.
Based on a maximum likelihood analysis, the half-light radius of Cetus~III
is estimated to be $\sim 90$ pc. This is significantly larger than MW
globular clusters with the same luminosity of $M_V \sim -2.5$ mag, thereby
suggesting that it is a dwarf galaxy. 
Deeper imaging of Cetus~III with space telescopes will be worthwhile for getting
further constraints from its faint RGB/MS members against background faint galaxies.
Also, spectroscopic follow-up studies of stars in Cetus~III will be important to
assess the nature of this stellar system as a dwarf satellite by constraining
their membership and also determine chemical abundance and stellar kinematics,
so that chemo-dynamical properties of this satellite may be derived
and compared to other satellites.

The heliocentric distance to Cetus~III is 251$^{+24}_{-11}$ kpc or
$(m-M)_0 = 22.0^{+0.2}_{-0.1}$~mag and its completeness-corrected,
absolute magnitude in the $V$ band is estimated as $M_V = -2.45^{+0.57}_{-0.56}$~mag.
This suggests that Cetus~III lies beyond the reach of the SDSS but inside the detection limit of
the HSC-SSP survey, for which we adopt the limiting $i$-band magnitude of 24.5 mag
below which star/galaxy classification becomes difficult. Thus, we expect the presence of
more satellites in the MW halo, which have not yet been identified because of their
faint luminosities and large distances. Based on the results of high-resolution
N-body simulations for the evolution of dark matter halos combined with
the abundance matching method to derive stellar masses, we have calculated
the luminosity function of visible satellites associated with numerous subhalos
in a MW-sized host halo to estimate the likely number of new dwarf satellites
to be discovered in the Subaru/HSC survey. Although this estimate suffers from
uncertainties mainly from the abundance matching method, we anticipate that
we will discover $9^{+15}_{-5}$ satellites with $M_V < 0$ in the HSC-SSP
survey volume. Therefore the completion of this survey program will provide important
insights into the missing satellites problem and thus the nature of dark matter
on small scales.

%%%%%%%%%%%%%%%%%%%%%%%%%%%%%%%%%%%%%%%%%%%%%%%%%%%%%%%%%
\begin{ack}
This work is based on data collected at the Subaru Telescope and retrieved from
the HSC data archive system, which is operated by Subaru Telescope and Astronomy
Data Center at National Astronomical Observatory of Japan.
MC acknowledges support in part from MEXT Grant-in-Aid for Scientific Research on 
Innovative Areas (No. 15H05889, 16H01086).

The Hyper Suprime-Cam (HSC) collaboration includes the astronomical communities
of Japan and Taiwan, and Princeton University. The HSC instrumentation and
software were developed by the National Astronomical Observatory of Japan (NAOJ),
the Kavli Institute for the Physics and Mathematics of the Universe (Kavli IPMU),
the University of Tokyo, the High Energy Accelerator Research Organization (KEK),
the Academia Sinica Institute for Astronomy and Astrophysics in Taiwan (ASIAA),
and Princeton University. Funding was contributed by the FIRST program from Japanese
Cabinet Office, the Ministry of Education, Culture, Sports, Science and Technology (MEXT),
the Japan Society for the Promotion of Science (JSPS), Japan Science and
Technology Agency (JST), the Toray Science Foundation, NAOJ, Kavli IPMU, KEK, ASIAA,
and Princeton University.  

This paper makes use of software developed for the Large Synoptic Survey Telescope.
We thank the LSST Project for making their code available
as free software at \url{http://dm.lsst.org}.

The Pan-STARRS1 Surveys (PS1) have been made possible through contributions of
the Institute for Astronomy, the University of Hawaii, the Pan-STARRS Project Office,
the Max-Planck Society and its participating institutes, the Max Planck Institute for Astronomy,
Heidelberg and the Max Planck Institute for Extraterrestrial Physics, Garching,
The Johns Hopkins University, Durham University, the University of Edinburgh,
Queen's University Belfast, the Harvard-Smithsonian Center for Astrophysics,
the Las Cumbres Observatory Global Telescope Network Incorporated,
the National Central University of Taiwan, the Space Telescope Science Institute,
the National Aeronautics and Space Administration under Grant No. NNX08AR22G issued through
the Planetary Science Division of the NASA Science Mission Directorate,
the National Science Foundation under Grant No. AST-1238877, the University of Maryland,
and Eotvos Lorand University (ELTE) and the Los Alamos National Laboratory.
\end{ack}

% reference

\end{document}